# Ethics for Digital Medicine: A Path for Ethical Emerging Medical IoT Design

**Sudeep Pasricha**
Colorado State University

*Abstract*—The dawn of the digital medicine era, ushered in by increasingly powerful embedded systems and Internet of Things (IoT) computing devices, is creating new therapies and biomedical solutions that promise to positively transform our quality of life. However, the digital medicine revolution also creates unforeseen and complex ethical, regulatory, and societal issues. In this article, we reflect on the ethical challenges facing digital medicine. We discuss the perils of ethical oversights in medical devices, and the role of professional codes and regulatory oversight towards the ethical design, deployment, and operation of digital medicine devices that safely and effectively meet the needs of patients. We advocate for an ensemble approach of intensive education, programmable ethical behaviors, and ethical analysis frameworks, to prevent mishaps and sustain ethical innovation, design, and lifecycle management of emerging digital medicine devices.

■ **Digital Medicine** refers to smart healthcare products that have a direct impact on diagnosing, preventing, monitoring, and treating a disease, condition or syndrome. These products make use of a wide range of embedded systems and Internet of Things (IoT) devices, e.g., as part of nanobots for precise delivery of drugs to specific organs with pinpoint accuracy, intelligent locomotive and sensory prosthetics, and augmented/virtual reality limb rehabilitation solutions for patients that have suffered from stroke, cerebral palsy, and traumatic brain injury. Many of the requirements of traditional medications are shared by digital medicine devices (DMDs), such as the need for approval from regulatory bodies before accessing the market. But there are also differences. Unlike standard medicines that primarily rely on new active principles, DMDs rely heavily on the design of innovative embedded and IoT computing platforms with custom form factors (e.g., nanoscale), unique computational and networking capabilities, and stringent power and energy budgets, that can be combined with traditional therapy or medication.

As digital medicine often plays a key role in matters of life and death, it is important to consider ethical implications in all aspects of the design, deployment, and usage lifecycle of its technological components. It is important here to make a distinction between ethics and laws. In our modern society, laws establish minimal requirements of behavior for individuals and social institutions to function well. They enable us to have orderly interactions with one another within a framework of basic fairness. But they are narrow in scope and say nothing, for example, about how life-critical decisions should be made in autonomous vehicles or how to tackle misinformation and deep fakes. Ethics goes beyond law to identify moral issues, reflecting on the nature of society we want to live in based on ideas of what a good life looks like. It makes recommendations that allow us to strive towards ways of life that will be most conducive to our own and others' wellbeing.

Ethical concerns are no stranger to medicine. After all, thousands of years ago, Hippocrates, considered the father of Western medicine, was urging physicians to do no harm. The discipline of medical ethics, however, did not fully develop until the 1970s. Two factors contributed to the development of medical ethics [1]. First, medical scandals emphasized the need for ethical standards and regulation. For instance, the Tuskegee Syphilis Experiment scandal in 1972 revealed that subjects with syphilis had gone untreated for four decades despite treatment being available, and without the knowledge or consent of subjects. Second, advances in technology confronted doctors with new ethical challenges they did not know how to resolve. The mechanical ventilator, for instance, forced a rethink of the concept of death and organ transplants. Doctors now had access to bodies without functioning brains but with working organs that could be transplanted. It became clear that addressing such ethical dilemmas required experts other than healthcare professionals whose primary expertise was in keeping people healthy, and not in resolving ethical problems.

Today, the biomedical field has come to prioritize four foundational ethical principles – justice (fair, equitable, and appropriate treatment of patients), respect for autonomy (relating to a patient's capacity to self-determine), beneficence (obligation to act for the benefit of the patient), and non-maleficence (avoiding harm to the patient). These principles help realize a common morality that must also be applied to emerging DMDs, during their design and development (including during clinical trials), as well as during post-deployment monitoring and troubleshooting (including product recalls and replacements). Fortunately, medical device development is already regulated, with codes, guidelines, and standards aimed at protecting patients [2]. However, the ethical framework within which these directives work remains unwritten and nebulous. Coupled with rapid advances in digital medicine technology and evolving societal values, engineers working on digital medicine products are increasingly facing gray areas not covered by codes and regulations.

Part of the challenge is that the fields of ethics and technology development have very distinct cultures [3]. For instance, while ethics often draws out discussion by way of analogies and thought experiments, the technology development field tends to make use of more direct, concrete cases. Where ethics is keen to critique theories and methods, technology engineers are more interested in usable results. And where the practice of ethics can take substantial time to conduct analyses, technological developments must minimize time and other expenditures. As engineers and scientists working on medical technologies are unlikely to become experts in ethical reasoning, there is a need to identify robust methodologies that can integrate ethical considerations across the DMD lifecycle.

The other challenge has to do with the complexity of multiple stakeholders in the DMD lifecycle. One may think that patients are the primary stakeholders with DMDs, but not always. For instance, a DMD designer should consider manufacturing engineers as stakeholders because design decisions can lead to the use of chemicals during manufacturing that may be hazardous to those engineers (and violate the principle of non-maleficence from their perspective). Of course this stakeholder has little relevance once a device is in the market. Or consider the choice of material early in the design phase of an implantable left ventricular assist device (LVAD). It may be beneficial to consider a liability lawyer as a stakeholder in the device post deployment phase, as instances of the material ending up being the source of adjacent tissue damage and infections could shift the balance away from true beneficence to society from the perspective of the liability lawyer, even though that material may have been one of the easiest to use in manufacture. Thus, identifying the appropriate stakeholders, of which there can be many, is crucial for effective ethical analysis during a DMD lifecycle.

This article reflects on the ethical challenges facing digital medicine. While some prior surveys have discussed ethics for health-related IoT devices, e.g., [13], their scope primarily pertains to technological challenges, e.g., creating secure and privacy-preserving hardware/software solutions. In contrast, our scope is more up to date and also broader, covering ethical challenges with digital medicine from a technological as well as a regulatory, professional, societal, and educational perspective.

PERILS OF DIGITAL MEDICINE

Shockingly, at the dawn of the digital medicine era, data from the U.S. Food & Drug Administration (FDA) indicates cause for alarm: more than 80,000 deaths and 1.7 million injuries have been linked to medical devices in the past decade [4]. Outside the U.S., detailed data is hard to come by, as governments in most countries in Africa, Asia and South America do not regulate medical devices at all.

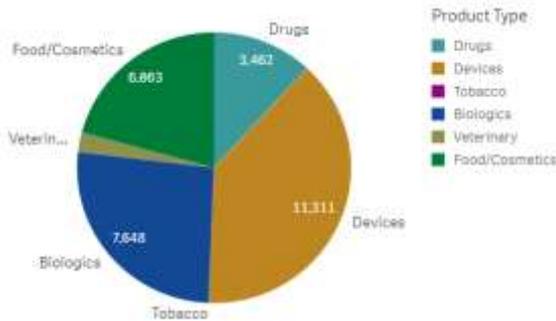

**Figure 1: US FDA recall events by product type, from 2012-2022. Source: FDA Data Dashboard**

Trends from available data indicate that there has been a marked rise in medical device mishaps and recalls in recent years [5]. Figure 1 shows how medical devices made up the largest portion of FDA recalls from 2012-2022. In February 2016, 263,520 units of glucose monitoring (CGM) systems were recalled due to a faulty auditory alarm. The CGMs included a sensor placed subcutaneously to measure blood glucose readings in patients, which were then sent to the receiver. The faulty alarms remained inactive in the defective CGM systems during high or low blood glucose levels in patients, potentially leading to serious adverse events and even death. In August 2017, 465,000 implantable cardiac pacemakers were recalled due to cyber security concerns. The faulty pacemakers implanted in the patients could potentially be hacked to alter the patients' heart rate. The implanted devices were eventually updated by medical staff with features including data encryption, operating system patches, and network connectivity disabling ability. Another large recall of pacemakers occurred in January 2019 and affected 156,957 units. The recalled pacemakers were found to be susceptibility to circuit failure when programmed to a dual chamber mode for sensing atrial activity. In June 2021, Philips recalled 3.5 million ventilation devices after finding a defect that could cause cancer. The ventilators used polyester-based polyurethane sound abatement foam, which had the potential to degrade into particles that could be ingested or inhaled and have toxic and carcinogenic effects.

As we begin to rely on digital medicine more heavily for our healthcare needs, change is urgently needed to alleviate the trend of increasing medical device related mishaps. Consider the example of implantable spinal-cord stimulators that use electrical currents to block pain signals before they reach the brain. An Associated Press investigation found that these devices account for the third-highest number of medical device injury reports to the U.S. FDA, with more than 80,000 incidents and hundreds of deaths flagged since 2008 [6]. Patients report that they have been shocked or burned or have suffered spinal-cord nerve damage ranging from muscle weakness to paraplegia. A class action lawsuit was filed in 2021. Could ethical analysis implemented early and often throughout the device lifecycle have helped avoid the many mishaps? Early design engineering teams could have focused on the principle of beneficence for patients and selected proven pulse generators and battery suppliers. They could also have sought perspectives of both patients and surgeons on what failure modes might be unacceptable. Further explorations of different design criteria, implantation techniques, or patient selection criteria could have helped mitigate some of these risks. With respect for patient autonomy, the design team could have required that the device be distributed with a more capable hand-held remote control and instructions that clearly displayed risks of certain settings. With a focus on justice for society, designers could have more extensively tested the device on a broader population of subjects in clinical trials. Considering the surgeons, the device manufacturers should have ensured that they are adequately trained in the nuances of the specific device (after reports of hundreds of deaths and injuries, the FDA had to remind physicians to follow product labels calling for a simulation in patients prior to permanently implanting the device). Lastly, the post-deployment response from the device manufacturers should have included periodic recalculation of risk versus benefit to patients and informed the design team about problems earlier, in addition to warning patients and surgeons more proactively. They could have also suggested to the clinical team, based on the principle of justice or beneficence, that a study be undertaken to better define the patient cohort likely to have the best outcomes. This example with the spinal-cord stimulator highlights one of the many cases in recent years where systemic ethical shortcomings across a medical device lifecycle led to injuries and deaths that could have been minimized if ethical considerations were prioritized and kept at the forefront.

## ROLE OF REGULATIONS AND CODES

Regulatory oversight remains an important incentive to make ethical decisions in medical device companies today. The FDA's Center for Devices and Radiological Health (CDRH) is responsible for regulating firms that manufacture, repackage, relabel, and import medical devices sold in the United States. CDRH classifies medical devices into Class I, II, and III, with regulatory control increasing from Class I to Class III. The device classification regulation defines the regulatory requirements for a general device type. Most Class I devices are exempt from Premarket Notification 510(k); most Class II devices require Premarket Notification 510(k); and most Class III devices require Premarket Approval.

Unfortunately, regulations alone remain insufficient for guiding ethical actions in complex scenarios across the medical device lifecycle. This is because regulations are very much reactive in nature, with an inertia to their enaction that may allow unsatisfactory conditions to persist for long periods of time. Several medical device mishaps in recent years have been attributed to regulatory loopholes and lax oversight from regulatory agencies, which has led to devices being poorly vetted before their release into the market and poorly monitored after the fact [4]. In many cases, the current 510(k) regulatory pathway allows devices to be made available for use with limited to no clinical evidence if manufacturers can demonstrate that their product is substantially equivalent to an existing, approved device.

Regulations are also primarily developed and enacted as responses to already-existing ethical challenges and are therefore unable to address many of the ethical challenges in the emerging digital medicine field. Shifting the focus of ethical responsibility from medical device companies to regulatory agencies is also not an effective and scalable strategy, given the mismatch between the growing number of medical devices being introduced into the market and shrinking regulatory agency budgets (at least in the U.S.). More often than not, regulations are treated as a nuisance by medical device companies, part of an archaic system of legal constraints that must be grudgingly accommodated, in the device lifecycle. Maximizing return on investment remains the primary driver, and along the way, if one can avoid upsetting the legal department and the general public, one may even consider that ethical work is taking place. This fallacy can lead to many unintended, harmful consequences.

Codes of ethics represent another pillar to prop up ethical actions in the medical device lifecycle. These codes are necessary to establish benchmarks for good practices and values, which is particularly important in helping those new to the profession to develop a moral professional compass. But the few codes of ethics that exist from professional societies, such as the Advanced Medical Technology Association (AdvaMed) and the Medical Device Manufacturers Association (MDMA), have been criticized as being checklists that fail to provide guidance on how to comprehensively navigate practical ethical dilemmas during the medical device lifecycle [7]. The reasons to comply with ethical codes are often weak, and easily overridden by reasons to deviate from them, e.g., due to economic pressures. Moreover, like regulations, ethical codes are prone to being drafted in reaction to past ethical concerns and fail to provide insights in contexts where new technologies and use cases uncover entirely new ethical dilemmas.

## ETHICAL ISSUES FOR DIGITAL MEDICINE

The digital medicine revolution is to a large extent a response to our society's desire for improving quality of life, with ubiquitous availability of healthcare services to support it. Many new digital medicine innovations are on the horizon, with a goal of helping improve the quality and effectiveness of care, control expenditure, reach underserved or vulnerable populations, and relieve overstretched healthcare services. These innovations are already starting to make an impact on diagnostics, assistive care, and patient monitoring for cardiovascular diseases, neurological disorders, and assisted living. For example, ingestible digital pills are being developed that consist of IoT sensors co-encapsulated with normal medicines to allow a reliable monitoring of medication-taking behavior (such behavior is noted to be sub-optimal among patients with chronic illnesses such as Type 2 diabetes, hypertension, Alzheimer's disease, and hepatitis C [8]) and the collection of data concerning other health-related lifestyle habits [9]. Embodied artificial intelligence (AI) devices that involve a physical presence are being actively designed, with embedded machine intelligence that can learn and adapt to each individual's unique behaviors and lifestyle requirements, e.g., as part of artificially intelligent robotic agents or smart prostheses [10]. Smart wearable and implantable devices are being devised to predict strokes (the second leading cause of death throughout the world) and epileptic seizures [11]. Tactile Internet (TI) innovations are being considered for performing surgical and precision in-vivo procedures remotely by specialists. Such procedures would rely on connected robots, augmented/virtual reality systems, and smart gloves with accurate haptic feedback. Smart tactile devices are also being designed for tremor suppression in patients with Parkinson's and for trauma rehabilitation [12].

Embedded and IoT platforms at the heart of these (and other) DMDs will require considering a range of ethical issues, beyond well-publicized concerns with security [13]. These issues represent points of ethical uncertainty or disagreement, within the development team, the literature, or wider society, regarding what course of action ought to be pursued or how an ethical concept should be understood in relation to a medical use case. Privacy is a major ethical concern with the rise of ubiquitous monitoring systems. This relates to one's ability to control access to one's own health data as well as to the notion that a person (including a bystander in a monitored environment) has the freedom not to be observed or to have their own personal space. Unforeseen misuses of collected data will also be a concern, e.g., genetic discrimination and body habitus screening by employers, sedentary lifestyle stratification by insurers, etc. Biases, either unconscious or conscious, encoded in the hardware and software artefacts within DMDs represent another challenge. Medical AI algorithms with questionable biases that interact directly with patients in various states of vulnerability, including reduced well-being and capacity, could threaten patient preference, safety, and privacy. Algorithmic biases have already led to serious mistakes in the medical domain, such as an algorithm affecting millions of patients that was found to exhibit significant racial bias [14]. Such mistakes will likely become even more commonplace with the wider adoption of AI algorithms in healthcare. Lack of transparency and explainability with these AI algorithms is also a serious shortcoming, preventing predictable, reproducible, and understandable universal behavior. Another societal challenge facing DMD designers will be cost containment, to balance reasonable costs for a medical device versus broad access of the device to patients in need.

While the classical medical ethics principles of justice, non-maleficence, autonomy, and beneficence are still important to guide biomedical research and clinical practice, it remains unclear how these principles can help overcome the complex ethical issues facing digital medicine. As a result of these issues, DMD designers will be tasked with many responsibilities beyond creating a functioning device. They will need to carefully consider the implications of data collected by their device and its misuse, as well as controls to ensure privacy. Assessments will be required to ensure freedom from unintended biases and reproducibility in algorithms used to process data. More broadly, it would be important for designers to analyze options that patients and healthcare professionals will have for capturing, modifying, deleting, disseminating, or even misinterpreting such data. Designers would also need to determine how best to train and educate device users and other stakeholders that might have access to such data. They will need to be involved in balancing marketing claims about what their device should do in terms of harms and benefits with actual device design. And they will also have to question the true beneficence of a device, beyond its economic viability for the company and innovative capabilities, if patients that need it cannot actually afford it.

In the absence of meaningful regulations for new use cases (and due to inevitable pushback from medical device companies to prevent arduous new regulations that often manifest as higher costs), and a lack of clear guidelines to address complex ethical dilemmas from professional codes of ethics, those developing digital medicine technologies will largely be left to translate high-level ethical principles as they see fit. But DMD designers come from varied disciplines and backgrounds that typically do not include systematic ethics training. Few trained ethicists currently work in healthcare device companies, and there is no established culture of practical exchange between these fields. As such, it is very likely that developers of emerging DMDs will not have the necessary competency to translate unfamiliar high-level ethics principles. This does not bode well for a future where DMDs are expected to be deployed as pervasively as IoT devices are today.

ENABLING ETHICAL DIGITAL MEDICINE

Proficiency in specifying and applying ethical principles to a wide range of real-world ethical issues facing DMD lifecycles will require a multi-pronged ensemble approach with 1) intensive ethical training, 2) programmable ethical behaviors and 3) a framework for ethical analysis over the entire DMD lifecycle. We discuss these in detail below.

## Ethics Education

In response to the bad publicity generated from medical device mishaps and recalls, efforts are being made to improve general awareness of ethical concerns for engineers and scientists involved in research and design of medical technologies. Many universities and research institutions now cover topics related to ethics in their technical curricula with the explicit purpose of raising ethical awareness and capacities for critical reasoning in developers, programmers, and engineers [15]. Many technology companies are implementing training modules on ethics for their employees, e.g., ethical foresight analysis (EFA) to educate designers and managers in predicting potential ethical issues and the consequences of specific technologies [16].

Ethics education for technical audiences often focus on an exploration of normative ethics, which in turn focuses on theories that provide general moral rules governing our behavior. There are three popular theories (deontology, teleology, and axiology) that are taught as a guide to resolve ethical dilemmas [17]. Deontology uses processes and rules to guide ethical decisions. These rules are considered absolute, meaning that the acts must be done regardless of the consequences, e.g., the truth must always be told, no matter the consequence. Further, in deontology, the intention of the action is as important as the action. If an individual performs an action that results in a good outcome, but does so with a negative intention, then the action is still unethical. Teleology emphasizes outcomes over the process. It is a results-oriented approach that defines ethical behavior by good or bad consequences. Ethical decisions are those that create the greatest good. The most common teleology approach is utilitarianism, which stresses the greatest good for the greatest number of individuals. Axiology is guided by the question, "If I were a good person, what would I do?" Key to this ethical orientation is the reliance on core ethical virtues, such as justice, moderation, courage, compassion, and loyalty to guide ethical actions. The focus in axiology is therefore on virtuous personal character when it comes to making decisions rather than universal rules or consequences.

These theories are widely reflected in different spheres of aspirational modern societies and in the codes and values of professional institutions. In educational settings, often case studies are used to highlight the appropriateness of theories based on the situation, as part of applied ethics analyses. Often, ethics education modules for engineers and scientists also end up covering practical tools for risk assessment, e.g., failure mode and effects analysis (FMEA), to evaluate the possible failure of a system and its impact on the system operation, and in some cases, systems engineering principles related to Life Cycle Assessment (LCA), which involves analysis to quantify and assess the consumption of resources and the environmental impacts associated with a product (or service) throughout its lifecycle.

These educational efforts are promising steps to make DMD companies and designers aware of the ethical landscape, but they come with their own challenges. Integrating ethical topics in curricula that are already packed with courses and with little wiggle room remains a challenge. Companies also have limited capacity to ensure adequate ethics training for their employees, given time-to-market pressures, attrition, and the proclivity to prioritize technical training to meet immediate and measurable design goals. It may be unrealistic to expect that curricula changes and training sessions alone will prepare every developer to be experts both in their technical domains and in ethics.

## Programming Ethical Behaviors

A concrete approach for ensuring that DMDs operate ethically is to program ethical behaviors as part of the hardware or software artefacts of the device. In its simplest form, this can involve encoding the required ethical behavior explicitly in rules or creating algorithms to allow devices to calculate the ethical actions. These rules can be based on ethical theories (e.g., deontology, teleology) and have an advantage in that they can be clearly understood by humans and can represent a range of the ethical behaviors required of the medical device. This approach can also be extended to allow for dynamic selection of rules based on situational context, and to provide device users the autonomy to make choices about ethical dilemmas, rather than have the reasoning hard coded by manufacturers. A drawback is that for complex ethical situations, it could be hard to even identify all possible situations for encoding into rules.

If AI algorithms are part of DMDs in any capacity, there is a critical need to mitigate bias and enable transparency. Programming ethical behaviors into AI algorithms must start with determining appropriate training datasets for them. If these datasets are biased, AI algorithm outcomes will also be biased. Examples of approaches to mitigate such dataset biases could include over/under-sampling, input weighing, and ensuring data collection diversity. Approaches to minimize biased outputs at the algorithm design level could include establishing thresholds for certain outputs and determining appropriate objective functions and benchmarks to test against. Another relevant direction is the area of "explainable AI" which is attempting to allow humans to understand the decisions made by AI. Examples of promising techniques from this field include layerwise relevance propagation (to determine which features in an input vector contribute most strongly to the output) and glass-box models (such as decision trees and Bayesian networks) that are more transparent to inspection. Allowing stakeholders to interpret and scrutinize how AI algorithms make decisions with such approaches can improve both accountability and troubleshooting (when undesired behaviors are detected) for AI-based DMDs.

Ideally, the task of identifying and integrating ethical capabilities into DMDs would be led by one or more ethicists that are part of the development process. As the development team may not be able to perceive every ethical issue, regular interactions between the ethicists and technically focused engineers would reduce the risk of ethical issues being overlooked or conflicts being glossed over. Without being overly prescriptive, ethicists could help explain and clarify complex ethical issues to allow a clearer understanding of them and use methods of ethical reasoning to justify or challenge a particular position or course of action. Such a collaborative ethics approach has been successfully employed in the field of genomics [18] and could greatly benefit the ethical design of emerging DMD solutions.

## Ethical Lifecycle Analysis

Considering the broader lifecycle of DMDs, there is a need for methodologies to support effective lifecycle-scale ethical analysis. This analysis should include identification of, engagement with, and explicit communication about the diverse values and perspectives of all stakeholders, while supporting systematic and thorough reflection and reasoning about the ethical issues. The reflection on ethical issues should go beyond teleological impacts and consequences of the device and include considerations at all stages of the device lifecycle, including the earliest stages of initial conceptual design and market analysis (to determine the ethics of the multiple pathways to innovation), design, validation, clinical trials, deployment, monitoring, repair, and retirement.

Developments in ethical product lifecycle analysis from the agricultural biotechnology field are particularly relevant for DMDs [19]. The Ethical Matrix method is a tool designed to evaluate the intersection of three normative ethical principles, (respect for well-being, autonomy, and justice) with four relevant stakeholder groups (the treated organisms, producers, consumers, and environment). Wright's Ethical Impact Assessment (EAI) tool is another mechanism to establish an analysis process based on questions for introducing consideration under five categories of issues defined by the four principles of biomedical principlism and the addition of privacy and data protection as a separate category. The method then suggests application of a series of other tools for ethics analysis as procedures for additional value appraisal, including the Ethical Matrix tool. The applicability to DMD development could be imagined with application of the same important principles to the relevant stakeholder groups of patients, medical device companies, surgeons, and hospitals, and concerns that encompass economic, regulatory, sustainability, and societal factors. To anticipate ethical issues for new technologies and use cases, the Anticipatory Technology Ethics (ATE) tool was proposed with mechanisms to forecast and deal with the inherent problem of uncertainty in technology development, especially during the research and early development stages [20]. The tool considers potential future uses and impacts with a broad scope.

It is unlikely that a single ethical analysis tool will be effective for a comprehensive assessment of the range of divergent ethical dilemmas involved in the introduction and use of new DMDs. Therefore, ethical analysis over DMD lifecycles should make use of a framework of multiple tools, including, but not limited, to the ones discussed above. Developers can be trained on the use of multiple tools in this framework and encouraged to select the most appropriate one(s) depending on the context.

CONCLUSION

It is not surprising that the medical device industry continues to experience significant and costly failures due not only to the external pressures of efficiency, access to resources, and insufficient analysis, but also often to insufficient approaches to ethical analysis. As we embark onto the next generation of smart and connected healthcare via digital medicine, mishaps with DMDs will impact a much larger proportion of the global population. We reviewed some of the key ethical challenges facing the design and use of embedded and IoT platforms in the context of emerging DMDs. We also discussed medical device mishaps in recent years and the inadequacy of relying on regulations and codes to achieve ethical DMD design. To effectively and comprehensively address ethical issues throughout the lifecycle of DMDs, we presented a multi-pronged, ensemble approach involving ethics-focused education, programming ethical behaviors, and a broader ethical analysis framework over the entire device lifecycle.

The proposed ensemble approach is also relevant in light of the recent COVID-19 pandemic. In the US, the most severe cases of COVID-19 have impacted Black and Latinx communities in disproportionate ways. This is partly because of long-standing disparities in health outcomes for these communities as well as disproportionate numbers of workers from Black and Latinx communities in job roles that cannot be filled at home or support adequate sick leave. Better reliance on ethics education as well as the ability to program ethical behaviors in AI algorithms (e.g., those being increasingly used for prioritizing vaccine distribution and allocation of scarce hospital resources), as advocated in the ensemble approach, can help emphasize the vital forms of social support for the most at-risk and vulnerable individuals in our society.

In the future, one can imagine that digital medicine will evolve to support a tighter, more symbiotic human-technology relationship, such as with brain augmentations that alter our behaviors and perhaps even the creation of digital consciousness that can be transferred to humanoid robots. In such high stakes and uncharted ethical landscapes, a culture of well-established ethical development and lifecycle management will become an imperative to allow humanity to continue to strive for a good life.

**Sudeep Pasricha** received his Ph.D. in Computer Science from UC Irvine in 2008. He is a Walter Scott Jr. College of Engineering Professor in the ECE Department and Director of the Embedded, High Performance, and Intelligent Computing (EPIC) Lab at Colorado State University. His research interests relate to embedded and IoT systems, with an emphasis on designing innovative software algorithms, hardware architectures, and hardware-software co-design techniques. He is a Senior Member of IEEE and Distinguished Member of ACM. Contact him at sudeep@colostate.edu.